\documentclass{optica-article}

\journal{opticajournal} % for journals or Optica Open

\articletype{Research Article}

\usepackage{lineno}
\begin{document}

\title{Structural characterization of thin-walled microbubble cavities}

\author{Mohammed Zia Jalaludeen,\authormark{1} Shilong Li,\authormark{1,*} Ke Tian,\authormark{1,2} Toshio Sasaki,\authormark{3} and Síle Nic Chormaic\authormark{1}}

\address{
\authormark{1}Light-Matter Interactions for Quantum Technologies Unit, Okinawa Institute of Science and Technology Graduate University, Onna, Okinawa 904-0495, Japan\\
\authormark{2}College of Physics and Optoelectronic Engineering, Harbin Engineering University, Harbin 150001, China\\
\authormark{3}Scientific Imaging Section, Okinawa Institute of Science and Technology Graduate University, Onna, Okinawa 904-0495, Japan
}

\email{\authormark{*}shilong.li@oist.jp} %% email address is required; see note below about the corresponding author designation

% use {asbstract*} to suppress the copyright line. Copyright information will be added in production

\begin{abstract*} 
Whispering gallery mode (WGM) microbubble cavities are a versatile optofluidic sensing platform owing to their hollow core geometry. To increase the light-matter interaction and, thereby, achieve a higher sensitivity, thin-walled microbubbles are desirable. However, a lack of knowledge about the precise geometry of hollow microbubbles prevents us from having an accurate theoretical model to describe the WGMs and their response to external stimuli. In this work, we provide a complete characterization of the wall structure of a microbubble and propose a theoretical model for the WGMs in this thin-walled microcavity based on the optical waveguide approach. Structural characterization of the wavelength-scale wall is enabled by focused ion beam milling and scanning electron microscopy imaging. The proposed theoretical model is verified by finite element method simulations. Our approach can readily be extended to other low-dimensional micro-/nanophotonic structures. 

\end{abstract*}

%%%%%%%%%%%%%%%%%%%%%%%%%%  body  %%%%%%%%%%%%%%%%%%%%%%%%%%

\section{Introduction}

Optical microcavities supporting whispering gallery modes (WGMs) have  been investigated intensively in the past two decades \cite{https://doi.org/10.1002/pssa.201900825,Yu2021} due to their ultrahigh quality factor (\textit{Q}-factor), which makes them suitable for various optical applications ranging from cavity
quantum electrodynamics \cite{Vahala2003,Spillane2005,Aoki2006Nature,Alton2011NP, PhysRevLett.126.233602} to label-free optical detection \cite{Armani2007Science,Zhu2010NP,Shao2013AM,Baaske2014NN,Kim:19}. Compared with the widely used WGM microcavity geometries such as microspheres \cite{doi:10.1063/1.2753591, Chiasera2010LPR}, microtoroids \cite{Armani2003Nature}, and microrings \cite{Rabiei2002}, microbubble cavities or microbubbles \cite{Sumetsky2010,Watkins:11} have the advantage of a hollow core and can be used as optofluidic devices in an all-fiber manner \cite{SenthilMurugan2011,Ward:18, doi:10.1021/acs.jpcc.2c05733}. Moreover, the resultant thin-walled structure of microbubbles provides us with new degrees of freedom, such as the thickness of the wall and its variation along the cavity axis. These allow us to engineer properties related to the WGMs, such as the mode field distribution, the mode dispersion, and the mode spectrum. Such engineered WGMs are particularly useful for various nonlinear optical processes, for example, four-wave parametric oscillation and frequency comb generation \cite{Yang2016,Yang2016_2,Kasumie:2018, Yin:19}. Therefore, an accurate determination of the geometry of a microbubble to precisely characterize its WGMs is an important prerequisite for practical applications of such cavities. 

Several methods to determine the thin-walled structure of microbubble cavities in a non-destructive way have already been reported. Bright-field microscopy is probably the simplest approach for measuring the diameter of a microbubble \cite{Henze2011}, but the low image contrast at a reasonable field-of-view excludes it as an effective way for a wall thickness measurement. Confocal microscopy has been used to measure the wall thickness of microbubble cavities \cite{Cosci2015}; however, the image resolution limits the accuracy of the measured thickness to half a wavelength. Obtaining the structural information of the microbubble by inferring its response to a certain stimulus seems to be a non-destructive method for the determination of the wall thickness. For example, the microbubble wall thickness was measured based on the internal aerostatic pressure sensing method with a measurement uncertainty on sub-micrometer scale \cite{Lu2016}. Nonetheless, such a method is not ideal as it requires precise knowledge of the structural information beforehand and it also assumes a constant wall thickness along the cavity axis.  
Currently, the only reliable method for studying the wall structure of a microbubble is a destructive approach, which involves breaking the microbubble and then measuring its cross-section using, for example, scanning electron microscopy \cite{Yang2016,Yang2016_3,PhysRevLett.124.103902}. However, a constant wall thickness along the cavity axis is generally assumed. A fully systematic study on the microbubble wall structure is yet to be carried out.

In this work, we fully characterize the wall structure of a microbubble cavity using focused ion beam (FIB) milling and scanning electron microscopy (SEM) imaging. Both the wall thickness and its variation along the cavity axis are obtained, thus enabling us to precisely model the microbubble geometry. Considering the wavelength-scale wall thickness, a theoretical model based on optical waveguide theory is proposed to describe the WGMs in the microbubble cavity. Finite element method simulations are performed to verify the validity of the proposed theoretical model. Our results will benefit not only the development of microbubble cavities but also the exploration of other low-dimensional micro-/nano-photonic structures. 

\begin{figure}
\hspace*{-0.7in}
\centering\includegraphics{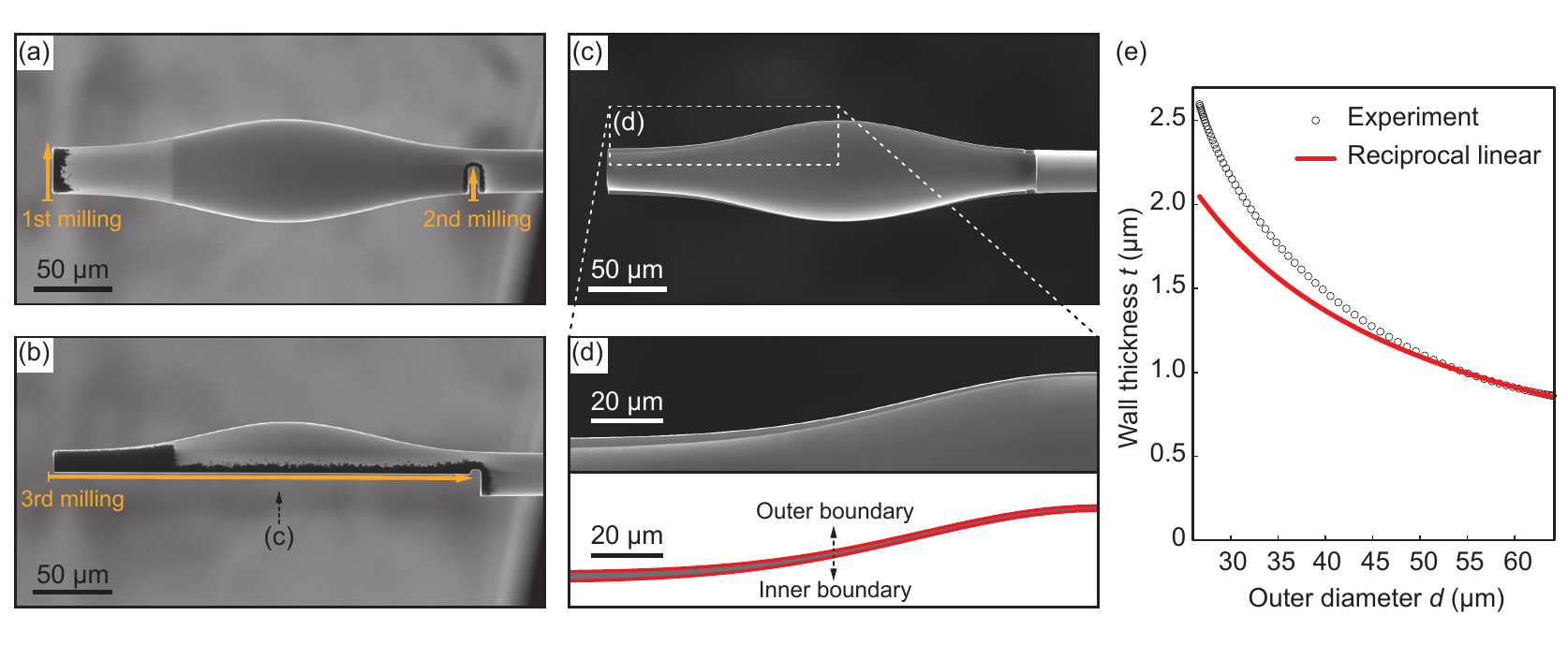}
\caption{\label{Figure_1}\textbf{Structural characterization of a thin-walled microbubble by FIB milling and SEM imaging}. The microbubble's support stem on the left side was initially removed through FIB milling, while half of the support stem on the right side was cut, creating a gap (a). A third FIB milling was conducted from the center of the left side towards the center of the gap on the right side, resulting in the removal of half of the microbubble (b). The wall structure of the microbubble is clearly visible under SEM imaging (c). Due to the high SEM imaging resolution, the wall thickness variation along the cavity axis can be determined with accuracy down to the nanometer scale (upper panel, d). To describe such a wall structure, Gaussian profiles were used to fit the outer and inner boundaries of the microbubble (lower panel, d). The dependence of the wall thickness $t$ on the outer diameter $d$ of the thin-walled microbubble is shown in (e). A reciprocal linear relation is satisfied near the center of the microbubble at larger outer diameters.}
\end{figure}

\section{Experimental results}

Silica microbubbles were fabricated as described previously \cite{Yang2016_3} using fused silica capillaries (360-$\mu$m outer diameter and 250-$\mu$m inner diameter) and a CO$_2$ laser. The wall structure of a microbubble (with a thin Au layer coating) was determined by means of FIB milling and SEM imaging (FEI Helios G3 UC), see Fig. \ref{Figure_1}. Due to the high imaging resolution of both the SEM and the FIB, the microbubble's wall structure was clearly visible. The largest diameter was around 64~$\mu$m at the bubble center, where the wall thickness was the thinnest ($\sim$0.85~$\mu$m). The bubble diameter gradually decreased, following a Gaussian, down to around 28~$\mu$m at the support stem, where the wall was thickest at  $\sim$2.60~$\mu$m. With this structural information, the mode structure of the thin-walled microbubble could be fully determined by either theoretical models or numerical simulations, as shown below. 

The wall thickness is a crucial parameter to measure the performance of microbubble-based optical sensors \cite{Yang2016_3}. However, due to the lack of an efficient approach to fully characterize the wall structure, geometrical approaches \cite{Henze2011,Cosci2015,Yu2020,Jiang2020} have widely been used to estimate the wall thickness. These models are based on two assumptions, i.e. constant wall thickness and mass conservation. The first assumption is meaningful when making order-of-magnitude estimates. The second assumption can be split into the following two scenarios: area conservation or volume conservation. In the first case, the microbubble is considered to be the result of cylindrical expansion of a capillary. The cross-section of the capillary is a ring with an area of $\pi (d/2)^2 - \pi (d/2 - t)^2\approx \pi d t$, where $d$ is the bubble's outer diameter and $t$  its wall thickness ($t\ll d$). The conserved area leads to a reciprocal linear relationship between $d$ and $t$, i.e., $t=c_1/d$ with a coefficient $c_1$. Similarly, for the second case, the microbubble can be viewed as the result of spherical expansion of a spherical bubble with the same diameter as the capillary. Since the volume of the spherical bubble is $(4/3)\pi (d/2)^3 - (4/3)\pi (d/2 - t)^3\approx \pi d^2 t$, a reciprocal quadratic relation of $t=c_2/d^2$ is obtained with $c_2$ as a coefficient. It is generally believed that the cylindrical expansion gives an over-estimation of the wall thickness, therefore the upper limit for the measured thickness, while the lower limit is obtained from the spherical expansion \cite{Cosci2015}.

Figure \ref{Figure_1}(e) shows the $t$ versus $d$ relationship for the thin-walled microbubble. The wall thickness of the microbubble varies along the bubble axis, clearly showing that it does not have a constant wall thickness, invalidating the first assumption mentioned above. The second assumption is also invalid because the relationship between $t$ and $d$ is neither reciprocal linear nor reciprocal quadratic. Nevertheless, around the center of the thin-walled microbubble ($\pm40$ $\mu$m from the microbubble center), a reciprocal linear relationship between $t$ and $d$ is satisfied. We attribute these seemingly unusual results to the fact that only the center of the microbubble was melted and fully expanded, while the portions near the support stems experienced lower temperatures and were unable to fully expand. These findings demonstrate the importance of developing an efficient approach for characterizing the wall structure of microbubbles. 

\section{Theoretical model}
\begin{figure}
\hspace*{-0.7in}
\centering\includegraphics{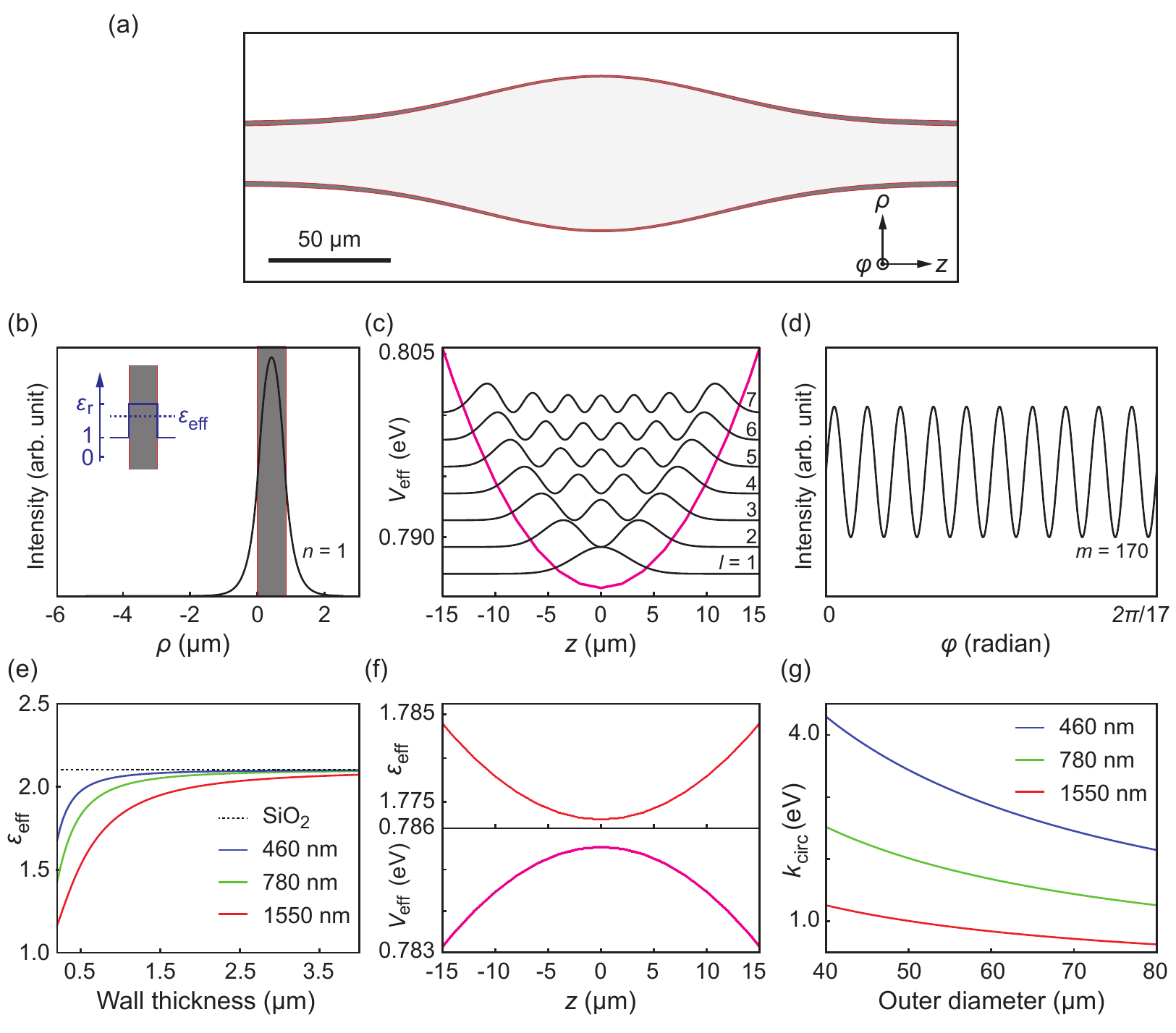}% Here is how to import EPS art
\caption{\label{Figure_2}\textbf{Theoretical model of a thin-walled microbubble}. (a) Reconstructed 3D geometry of the microbubble from the SEM images shown in Fig. \ref{Figure_1}. Due to the geometrical symmetries, the electric field distribution of WGMs in the microbubble can be calculated by solving the three coupled differential equations, Eqs. \ref{Equation_3}--\ref{Equation_5}. Three mode indices, i.e. $n$, $m$, and $l$, are resolved in the end to identify the WGMs, in addition to their polarization state (either TM or TE). (b) Radial field distribution. Owing to the wavelength-scale wall thickness, only the fundamental radial mode with $n = 1$ is considered. The inset illustrates the physical meaning of the effective permittivity $\varepsilon_{\mathrm{eff}}$. (c) Axial field distribution. The quasipotential of the microbubble forms a quantum well---due mainly to the diameter variation---which confines the axial motion such that different axial modes emerge. (d) Azimuthal field distribution. Only a small portion of the azimuthal field is shown for better visibility. (e) Dependence of $\varepsilon_{\mathrm{eff}}$ on the wall thickness. (f) Illusory example showing a quantum-barrier-like quasipotential (lower panel) formed solely by the wall thickness variation (upper panel) where a constant outer diameter was used. (g) Dependence of $k_{\mathrm{circ}}$ on the outer diameter.}
\end{figure}

As a type of optical bottle microresonators (BMRs), light in microbubbles is trapped in the cross-sectional plane, circulating around the bubble axis, while confined axially, bouncing back and forth between two turning points known as caustics; this is similar to the way charged particles are trapped in magnetic bottles \cite{Sumetsky2004}. Such a confinement of light in three dimensions (3D) results in the quantization of optical fields into a series of optical modes. With the structural information (Fig.~\ref{Figure_2}(a)) obtained from experimental measurements, the mode structure of the thin-walled microbubble can be theoretically modeled by simplifying Maxwell's equations. 
%The resulting theoretical model will be greatly simplified due to the micro-/nanometer-thin wall thickness of the microbubble, where the optical waveguide theory is applicable. 

For azimuthally and axially symmetric microbubbles made of isotropic and homogeneous nonmagnetic dielectric materials, optical spin-orbit coupling is absent \cite{Ma2016NC,NP2023}. Furthermore, most of these microbubbles have a relatively small diameter variation along the axial direction (see Fig. \ref{Figure_2}(a)). Therefore, the two sets of polarization modes can be well separated: the transverse electric (TE) modes with a nonzero axial electric field and the equivalent transverse magnetic (TM) modes with a nonzero axial magnetic field. Since the TE modes preferentially exist in thin-walled microbubbles, only they will be considered in the subsequent analysis. The Helmholtz equation for the nonzero axial electric field $E_z$ of TE modes reads:
\begin{equation}
\nabla^2 E_z(\mathbf{r})+k^2\mathbf{\varepsilon}_\mathrm{r}(\mathbf{r}) E_z(\mathbf{r}) = 0.
\label{Equation_1}
\end{equation}
Here, $k=\omega/c$ is the wave vector, $\omega$ is the angular frequency, $c=1/\sqrt{\varepsilon_0 \mu_0}$ is the speed of light, and $\varepsilon_0$ ($\mu_0$) is the permittivity (permeability) in vacuum. The relative permittivity $\mathbf{\varepsilon}_\mathrm{r}=\mathbf{\varepsilon}/\varepsilon_0$, where $\mathbf{\varepsilon}$ is the permittivity of a material. The relative permeability $\mathbf{\mu}_\mathrm{r}$ of nonmagnetic materials in the visible and infrared spectral range is close to unity and one may set the material's permeability $\mathbf{\mu} = \mu_0$. 

The scalar Helmholtz equation for $E_z$ is a 3D partial differential equation. It can be further simplified by the method of separation of variables in the cylindrical coordinates $(\rho,\varphi,z)$ which are defined in Fig. \ref{Figure_2}(a). To this end, $E_z$ can be expressed in the separable form: 
\begin{equation}
E_z(\rho,\varphi,z) = P(\rho)\Phi(\varphi)\Psi(z),
\label{Equation_2}
\end{equation}
where $P(\rho)$, $\Phi(\varphi)$, and $\Psi(z)$ are the radial, azimuthal, and axial field components, respectively. Substituting Eq. \ref{Equation_2} into Eq. \ref{Equation_1}, three ordinary differential equations for the respective field components are obtained: 
%\begin{widetext}
\begin{eqnarray}
\left[\rho^2\frac{\mathrm{d}^2}{\mathrm{d}\rho^2}+\rho\frac{\mathrm{d}}{\mathrm{d}\rho}+\varepsilon_{\mathrm{r},z}(\rho)\rho^2 \right]P(\rho) =&& \varepsilon_{\mathrm{eff,}z} P(\rho),
\label{Equation_3}
\\
\frac{\mathrm{d}^2}{\mathrm{d}\varphi^2} \Phi(\varphi) =&& -k^2_{\mathrm{circ,}z}\varepsilon_{\mathrm{eff,}z} \Phi(\varphi),
\label{Equation_4}
\\
\left[-\frac{1}{\varepsilon_{\mathrm{eff,}z}}\frac{\mathrm{d}^2}{\mathrm{d}z^2}+k^2_{\mathrm{circ,}z}\right] \Psi(z) =&& k^2\Psi(z),
\label{Equation_5}
\end{eqnarray}
%\end{widetext}
which are coupled via two coupling constants $\varepsilon_{\mathrm{eff}}$ and $k_{\mathrm{circ}}$. Here, $\varepsilon_{\mathrm{eff}}$ couples the dynamics of the radial and the circular components of the light propagation, while $k_{\mathrm{circ}}$ couples the dynamics of the lateral and the axial components of the light propagation. In the end, the discrete spectrum $\omega=kc$ can be obtained by solving these three ordinary differential equations, Eqs. \ref{Equation_3}--\ref{Equation_5}. 
  
The general solution of Eq. \ref{Equation_3} consists of a linear combination of Bessel functions of the first and second kind. Their coefficients are determined by matching the wall boundary condition (Fig. \ref{Figure_2}(b)). The Bessel functions and, therefore, $P(\rho)$ look like oscillating sine or cosine functions that decay proportionally to $1/\sqrt{\rho}$. Different radial modes can  be distinguished by the number of `peaks' in the $P(\rho)$, with each mode labeled by a unique radial mode index $n=1,2,...$. Equation \ref{Equation_3} quantifies the effect of the wall thickness on these radial modes by using an effective permittivity $\varepsilon_{\mathrm{eff}}$, i.e., the first coupling constant, which measures the degree of light confinement by the wall (see the inset of Fig. \ref{Figure_2}(b)). Since the diameter of microbubbles is typically in the range of a few tens of micrometers, the weakly curved condition is satisfied, and the microbubble wall can be treated as a slab waveguide. Then, $\varepsilon_{\mathrm{eff}}$ can be easily found via the transcendental equation based on the well-established optical waveguide theory \cite{Strelow2012PRB}: 
\begin{equation}
\mathrm{tan}(\kappa t) = \frac{\kappa\gamma_{\mathrm{air}}+\kappa\gamma_{\mathrm{core}}}{\kappa^2-\gamma_{\mathrm{air}}\gamma_{\mathrm{core}}},
 \label{Equation_6}
\end{equation}
where $\gamma_{\mathrm{air}}^2=k^2\varepsilon_{\mathrm{eff}} - k^2\varepsilon_{\mathrm{air}}$, $\gamma_{\mathrm{core}}^2=k^2\varepsilon_{\mathrm{eff}} - k^2\varepsilon_{\mathrm{core}}$, $\kappa^2=k^2\varepsilon_\mathrm{r} - k^2\varepsilon_{\mathrm{eff}}$ with $\varepsilon_{\mathrm{air}}$ and $\varepsilon_{\mathrm{core}}$ are the relative permittivities of the surrounding air and the microbubble's core material, respectively. Figure \ref{Figure_2}(e) shows the calculated $\varepsilon_{\mathrm{eff}}$ as a function of the wall thickness at a few wavelengths of interest. It is clear that the microbubble wall plays a crucial role when its thickness is close to the propagating wavelength, as is the case for most microbubble cavities. 

The general solution of Eq. \ref{Equation_4} is $\Phi(\varphi)=A\mathrm{exp}[\pm i(\beta\varphi + \varphi_0)]$ with the amplitude $A$ and the initial phase $\varphi_0$, where $\beta=k_\mathrm{circ}\sqrt{\varepsilon_{\mathrm{eff}}}$ can be called the propagation constant. Note that `$\pm$' corresponds to the clockwise (CW) and counter-clockwise (CCW) modes \cite{Zhu2010NP}. To have a stable optical field distribution, the solution must satisfy the periodic boundary condition (Fig. \ref{Figure_2}(d)): 
\begin{equation}
\sqrt{\varepsilon_{\mathrm{eff},z}} \pi d = m \lambda_\mathrm{circ,z},
\label{Equation_7}
\end{equation}
where $\lambda_\mathrm{circ}=2\pi/k_\mathrm{circ}$ is the wavelength in the cross-sectional plane. On the one hand, it leads to the WGMs identified by the azimuthal mode index $m = 0,1,2,...$. On the other hand, it determines the $k_\mathrm{circ}$ in this cross-section, i.e., the second coupling constant. Figure \ref{Figure_2}(g) shows the dependence of  $k_{\mathrm{circ}}$ on the outer diameter. Generally speaking, the smaller the outer diameter, the stronger the azimuthal optical confinement. 

The last equation, Eq. \ref{Equation_5}, describes the axial dynamics of the WGMs. It is a quasi-Schr\"{o}dinger equation with the quasipotential: 
\begin{equation}
V_{\mathrm{eff}}(z) = \frac{\hbar c}{e}k_{\mathrm{circ,}z},
\label{Equation_8}
\end{equation}
where the elementary charge $e$ is used to scale the quasipotential in eV. Although not explicitly mentioned in the literature, the quasipotential in most microbubble cavities forms a quantum well, which confines the axial motion of the WGMs  (Fig. \ref{Figure_2}(c)). The quantization of the axial motion in the quantum well results in different axial modes indicated by the axial mode index $l=1,2,...$. However, due to the lack of analytical solutions for most quantum well quasipotentials, the finite difference method is widely employed as a reliable numerical approximation technique to solve the quasi-Schr\"{o}dinger equation \cite{Strelow2012PRB}. 

The customizability of their axial modes distinguishes microbottle cavities from  microsphere and microtoroid cavities \cite{SumetskyReview}. As one type of microbottle cavity, microbubbles provide a new degree-of-freedom, i.e., the wall structure, in tailoring the axial modes. This becomes clearer with the theoretical model presented here: both the wall thickness and its variation contribute to the axial quasipotential in Eq. \ref{Equation_8} by determining $\varepsilon_{\mathrm{eff}}$ through Eq. \ref{Equation_6} and subsequently influencing the value of $k_\mathrm{circ}$ as described in Eq. \ref{Equation_7}. Figure \ref{Figure_2}(f) shows an example where a microbubble can provide a quantum barrier for the axial optical motion if the quasipotential is formed solely by the wall structure. This is similar to the axial mode engineering in rolled-up microbottle cavities \cite{Strelow2008PRL,Strelow2012PRB,Li2013PRA,Fang2016,Tian2018}. 

\section{Simulation verification}

\begin{figure}
\hspace*{-0.35in}
\includegraphics{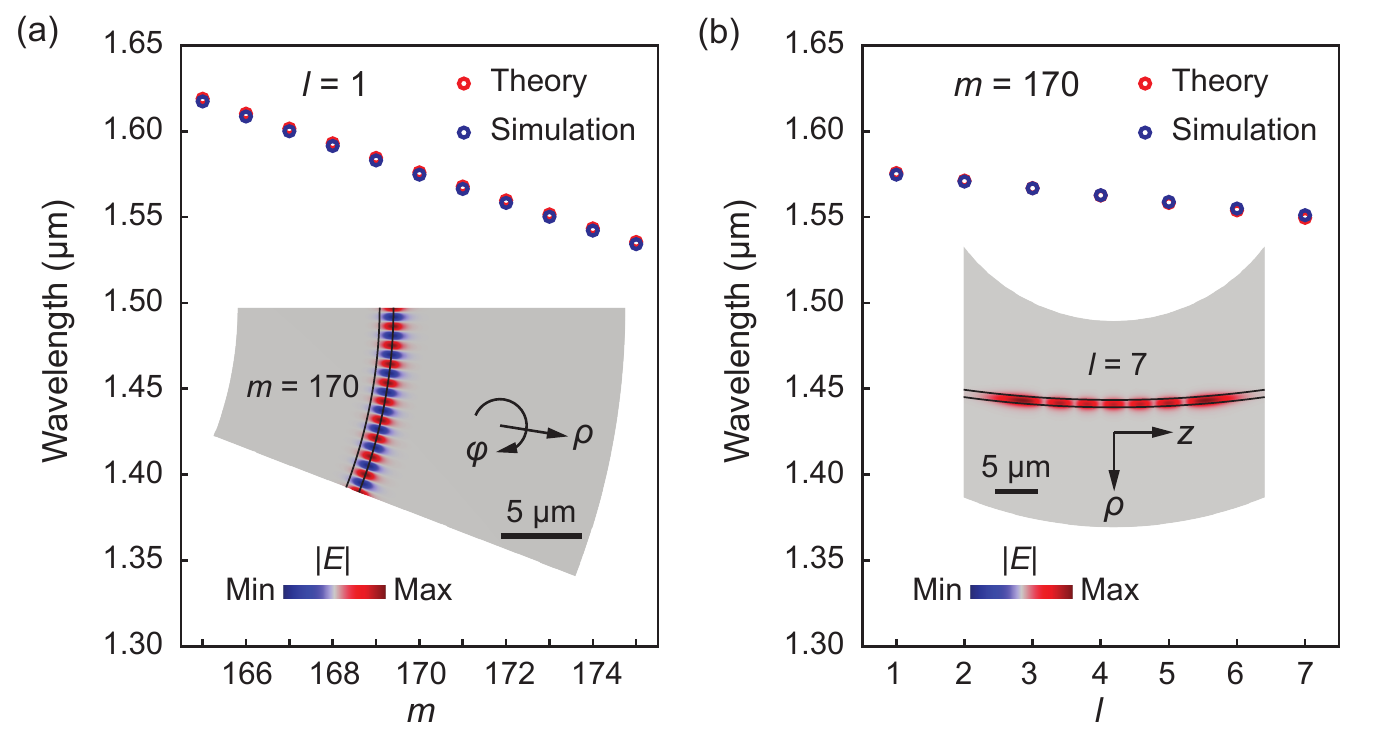}% Here is how to import EPS art
\caption{\label{Figure_3}\textbf{Verification of the theoretical model of the thin-walled microbubble by finite element method simulations}. (a) Resonant wavelengths for different azimuthal modes. Simulated field distribution in the cross-section is shown in the inset. (b) Resonant wavelengths for different axial modes. The inset shows the axial field distribution. $n=1$ for all cases.}
\end{figure}

The measured 3D structure of the thin-walled microbubble (Fig. \ref{Figure_2}(a)) allows us to characterise the shape with great accuracy. Therefore, a series of simulations based on the finite element method were carried out and the results are summarized in Fig. \ref{Figure_3}. These simulation results were used to verify the validity of the proposed theoretical model. 

Figure \ref{Figure_3}(a) shows the resonant wavelengths of WGMs with $m$ ranging from 165 to 175 ($l=1$ and $n=1$). The calculated values using the aforementioned theoretical model agree with the simulation results over a wide spectral range. Such good agreement confirms the validity of the proposed theoretical model based on the optical waveguide approximation. This deepens our understanding of the underlying physics and facilitates the design for device applications using thin-walled microbubble cavities. 

The comparison for the axial modes is made in Fig. \ref{Figure_3}(b) for WGMs with $m=170$ and $n=1$. 
Very good agreement between the theoretical model and simulations is also obtained. This confirms the effectiveness of treating the WGM axial dynamics in the same way as the dynamics of a particle in a quantum well. By doing so, thin-walled microbubble cavities  become a reliable experimental platform to test  quantum mechanics. On the other hand, the well-developed quantum  theory can be used to engineer the axial modes. 

\section{Conclusion}
We have demonstrated an efficient way to fully characterize the wall structure of a microbubble cavity. The 3D geometry of the microbubble was reconstructed based on FIB milling and SEM imaging. Owing to the wavelength-scale wall thickness, a theoretical model based on the optical waveguide approximation has been proposed to describe the WGMs in the thin-walled microbubble cavity. Simulations have also been performed using the fabricated microbubble structure. Very good agreement between the proposed theoretical model and simulations has been obtained, verifying the validity of the proposed theory. The demonstrated characterization and modeling approaches are readily adaptable for other wavelength-scaled photonic devices. 

%\section{Backmatter}
%Backmatter sections should be listed in the order Funding/Acknowledgment/Disclosures/Data Availability Statement/Supplemental Document section. An example of backmatter with each of these sections included is shown below.

\begin{backmatter}
\bmsection{Funding}
%Content in the funding section will be generated entirely from details submitted to Prism. Authors may add placeholder text in the manuscript to assess length, but any text added to this section in the manuscript will be replaced during production and will display official funder names along with any grant numbers provided. If additional details about a funder are required, they may be added to the Acknowledgments, even if this duplicates information in the funding section. See the example below in Acknowledgements. For preprint submissions, please include funder names and grant numbers in the manuscript.
This work was funded by the Okinawa Institute of Science and Technology Graduate University (OIST), and the Japan Society for the Promotion of Science (JSPS) KAKENHI through Grant-in-Aid for Scientific Research (C) with Grant Number 23K04617 and  Grant-in-Aid for Early-Career Scientists with Grant Number 22K14621. 

\bmsection{Acknowledgments}
The authors would like to thank the Engineering Section, the Scientific Computing \& Data Analysis Section, and the Scientific Imaging Section of Okinawa Institute of Science and Technology Graduate University (OIST) for technical assistance. 

\bmsection{Disclosures}
The authors declare no conflicts of interest.

\bmsection{Data Availability}
Data underlying the results presented in this paper are not publicly available at this time but may be obtained from the authors upon reasonable request.

\end{backmatter}

%%%%%%%%%%%%%%%%%%%%%%% References %%%%%%%%%%%%%%%%%%%%%%%%%

%%%%%%%%%% If using BibTeX:
\bibliography{References}

\begin{thebibliography}{10}
\newcommand{\enquote}[1]{``#1''}

\bibitem{https://doi.org/10.1002/pssa.201900825}
L.~Cai, J.~Pan, Y.~Zhao, J.~Wang, and S.~Xiao, \enquote{Whispering gallery mode
  optical microresonators: Structures and sensing applications,}
  {\protect\JournalTitle{Phys. Status Solidi A}} \textbf{217}, 1900825 (2020).

\bibitem{Yu2021}
D.~Yu, M.~Humar, K.~Meserve, R.~C. Bailey, S.~Nic~Chormaic, and F.~Vollmer,
  \enquote{Whispering-gallery-mode sensors for biological and physical
  sensing,} {\protect\JournalTitle{Nat. Rev. Methods Primers}} \textbf{1}, 82
  (2021).

\bibitem{Vahala2003}
K.~J. Vahala, \enquote{Optical microcavities,} {\protect\JournalTitle{Nature
  (London)}} \textbf{424}, 839--846 (2003).

\bibitem{Spillane2005}
S.~M. Spillane, T.~J. Kippenberg, K.~J. Vahala, K.~W. Goh, E.~Wilcut, and H.~J.
  Kimble, \enquote{{Ultrahigh-Q toroidal microresonators for cavity quantum
  electrodynamics},} {\protect\JournalTitle{Phys. Rev. A}} \textbf{71}, 013817
  (2005).

\bibitem{Aoki2006Nature}
T.~Aoki, B.~Dayan, E.~Wilcut, W.~P. Bowen, A.~S. Parkins, T.~J. Kippenberg,
  K.~J. Vahala, and H.~J. Kimble, \enquote{Observation of strong coupling
  between one atom and a monolithic microresonator,}
  {\protect\JournalTitle{Nature (London)}} \textbf{443}, 671--674 (2006).

\bibitem{Alton2011NP}
D.~J. Alton, N.~P. Stern, T.~Aoki, H.~Lee, E.~Ostby, K.~J. Vahala, and H.~J.
  Kimble, \enquote{Strong interactions of single atoms and photons near a
  dielectric boundary,} {\protect\JournalTitle{Nat. Phys.}} \textbf{7},
  159--165 (2011).

\bibitem{PhysRevLett.126.233602}
E.~Will, L.~Masters, A.~Rauschenbeutel, M.~Scheucher, and J.~Volz,
  \enquote{Coupling a single trapped atom to a whispering-gallery-mode
  microresonator,} {\protect\JournalTitle{Phys. Rev. Lett.}} \textbf{126},
  233602 (2021).

\bibitem{Armani2007Science}
A.~M. Armani, R.~P. Kulkarni, S.~E. Fraser, R.~C. Flagan, and K.~J. Vahala,
  \enquote{Label-free, single-molecule detection with optical microcavities,}
  {\protect\JournalTitle{Science}} \textbf{317}, 783--787 (2007).

\bibitem{Zhu2010NP}
J.~Zhu, S.~K. Ozdemir, Y.-F. Xiao, L.~Li, L.~He, D.-R. Chen, and L.~Yang,
  \enquote{{On-chip single nanoparticle detection and sizing by mode splitting
  in an ultrahigh-Q microresonator},} {\protect\JournalTitle{Nat. Photonics}}
  \textbf{4}, 46--49 (2010).

\bibitem{Shao2013AM}
L.~Shao, X.-F. Jiang, X.-C. Yu, B.-B. Li, W.~R. Clements, F.~Vollmer, W.~Wang,
  Y.-F. Xiao, and Q.~Gong, \enquote{Detection of single nanoparticles and
  lentiviruses using microcavity resonance broadening,}
  {\protect\JournalTitle{Adv. Mater.}} \textbf{25}, 5616--5620 (2013).

\bibitem{Baaske2014NN}
M.~D. Baaske, M.~R. Foreman, and F.~Vollmer, \enquote{Single-molecule nucleic
  acid interactions monitored on a label-free microcavity biosensor platform,}
  {\protect\JournalTitle{Nat. Nanotechnol.}} \textbf{9}, 933--939 (2014).

\bibitem{Kim:19}
Y.~Kim and H.~Lee, \enquote{On-chip label-free biosensing based on active
  whispering gallery mode resonators pumped by a light-emitting diode,}
  {\protect\JournalTitle{Opt. Express}} \textbf{27}, 34405--34415 (2019).

\bibitem{doi:10.1063/1.2753591}
J.~M. Ward, D.~G. O’Shea, B.~J. Shortt, and S.~{Nic Chormaic},
  \enquote{Optical bistability in {Er-Yb} codoped phosphate glass microspheres
  at room temperature,} {\protect\JournalTitle{J. Appl. Phys.}} \textbf{102},
  023104 (2007).

\bibitem{Chiasera2010LPR}
A.~Chiasera, Y.~Dumeige, P.~Féron, M.~Ferrari, Y.~Jestin, G.~Nunzi~Conti,
  S.~Pelli, S.~Soria, and G.~C. Righini, \enquote{Spherical
  whispering-gallery-mode microresonators,} {\protect\JournalTitle{Laser
  Photonics Rev.}} \textbf{4}, 457--482 (2010).

\bibitem{Armani2003Nature}
D.~K. Armani, T.~J. Kippenberg, S.~M. Spillane, and K.~J. Vahala,
  \enquote{{Ultra-high-Q toroid microcavity on a chip},}
  {\protect\JournalTitle{Nature (London)}} \textbf{421}, 925--928 (2003).

\bibitem{Rabiei2002}
P.~Rabiei, W.~H. Steier, C.~Zhang, and L.~R. Dalton, \enquote{Polymer
  micro-ring filters and modulators,} {\protect\JournalTitle{J. Lightwave
  Technol.}} \textbf{20}, 1968--1975 (2002).

\bibitem{Sumetsky2010}
M.~Sumetsky, Y.~Dulashko, and R.~S. Windeler, \enquote{Optical microbubble
  resonator,} {\protect\JournalTitle{Opt. Lett.}} \textbf{35}, 898--900 (2010).

\bibitem{Watkins:11}
A.~Watkins, J.~Ward, Y.~Wu, and S.~{Nic Chormaic}, \enquote{Single-input
  spherical microbubble resonator,} {\protect\JournalTitle{Opt. Lett.}}
  \textbf{36}, 2113--2115 (2011).

\bibitem{SenthilMurugan2011}
G.~S. Murugan, M.~N. Petrovich, Y.~Jung, J.~S. Wilkinson, and M.~N. Zervas,
  \enquote{Hollow-bottle optical microresonators,} {\protect\JournalTitle{Opt.
  Express}} \textbf{19}, 20773--20784 (2011).

\bibitem{Ward:18}
J.~M. Ward, Y.~Yang, F.~Lei, X.~C. Yu, Y.~F. Xiao, and S.~{Nic Chormaic},
  \enquote{Nanoparticle sensing beyond evanescent field interaction with a
  quasi-droplet microcavity,} {\protect\JournalTitle{Optica}} \textbf{5},
  674--677 (2018).

\bibitem{doi:10.1021/acs.jpcc.2c05733}
F.~Pan, K.~Karlsson, A.~G. Nixon, L.~T. Hogan, J.~M. Ward, K.~C. Smith, D.~J.
  Masiello, S.~{Nic Chormaic}, and R.~H. Goldsmith, \enquote{Active control of
  plasmonic–photonic interactions in a microbubble cavity,}
  {\protect\JournalTitle{J. Phys. Chem. C}} \textbf{126}, 20470--20479 (2022).

\bibitem{Yang2016}
Y.~Yang, X.~Jiang, S.~Kasumie, G.~Zhao, L.~Xu, J.~M. Ward, L.~Yang, and
  S.~Nic~Chormaic, \enquote{Four-wave mixing parametric oscillation and
  frequency comb generation at visible wavelengths in a silica microbubble
  resonator,} {\protect\JournalTitle{Opt. Lett.}} \textbf{41}, 5266--5269
  (2016).

\bibitem{Yang2016_2}
Y.~Yang, Y.~Ooka, R.~M. Thompson, J.~M. Ward, and S.~Nic~Chormaic,
  \enquote{Degenerate four-wave mixing in a silica hollow bottle-like
  microresonator,} {\protect\JournalTitle{Opt. Lett.}} \textbf{41}, 575--578
  (2016).

\bibitem{Kasumie:2018}
S.~Kasumie, Y.~Yang, J.~M. Ward, and S.~{Nic Chormaic}, \enquote{{Towards
  visible frequency comb generation using a hollow WGM resonator},}
  {\protect\JournalTitle{Rev. Laser Engin.}} \textbf{46}, 92--96 (2018).

\bibitem{Yin:19}
Y.~Yin, Y.~Niu, H.~Qin, and M.~Ding, \enquote{Kerr frequency comb generation in
  microbottle resonator with tunable zero dispersion wavelength,}
  {\protect\JournalTitle{J. Lightwave Technol.}} \textbf{37}, 5571--5575
  (2019).

\bibitem{Henze2011}
R.~Henze, T.~Seifert, J.~Ward, and O.~Benson, \enquote{Tuning whispering
  gallery modes using internal aerostatic pressure,}
  {\protect\JournalTitle{Opt. Lett.}} \textbf{36}, 4536--4538 (2011).

\bibitem{Cosci2015}
A.~Cosci, F.~Quercioli, D.~Farnesi, S.~Berneschi, A.~Giannetti, F.~Cosi,
  A.~Barucci, G.~Nunzi~Conti, G.~Righini, and S.~Pelli, \enquote{Confocal
  reflectance microscopy for determination of microbubble resonator thickness,}
  {\protect\JournalTitle{Opt. Express}} \textbf{23}, 16693--16701 (2015).

\bibitem{Lu2016}
Q.~Lu, J.~Liao, S.~Liu, X.~Wu, L.~Liu, and L.~Xu, \enquote{Precise measurement
  of micro bubble resonator thickness by internal aerostatic pressure sensing,}
  {\protect\JournalTitle{Opt. Express}} \textbf{24}, 20855--20861 (2016).

\bibitem{Yang2016_3}
Y.~Yang, S.~Saurabh, J.~M. Ward, and S.~Nic~Chormaic, \enquote{{High-Q,
  ultrathin-walled microbubble resonator for aerostatic pressure sensing},}
  {\protect\JournalTitle{Opt. Express}} \textbf{24}, 294--299 (2016).

\bibitem{PhysRevLett.124.103902}
F.~Lei, J.~M. Ward, P.~Romagnoli, and S.~{Nic Chormaic},
  \enquote{Polarization-controlled cavity input-output relations,}
  {\protect\JournalTitle{Phys. Rev. Lett.}} \textbf{124}, 103902 (2020).

\bibitem{Yu2020}
J.~Yu, J.~Zhang, R.~Wang, A.~Li, M.~Zhang, S.~Wang, P.~Wang, J.~M. Ward, and
  S.~Nic~Chormaic, \enquote{A tellurite glass optical microbubble resonator,}
  {\protect\JournalTitle{Opt. Express}} \textbf{28}, 32858--32868 (2020).

\bibitem{Jiang2020}
J.~Jiang, Y.~Liu, K.~Liu, S.~Wang, Z.~Ma, Y.~Zhang, P.~Niu, L.~Shen, and
  T.~Liu, \enquote{{Wall-thickness-controlled microbubble fabrication for
  WGM-based application},} {\protect\JournalTitle{Appl. Opt.}} \textbf{59},
  5052--5057 (2020).

\bibitem{Sumetsky2004}
M.~Sumetsky, \enquote{Whispering-gallery-bottle microcavities: The
  three-dimensional etalon,} {\protect\JournalTitle{Opt. Lett.}} \textbf{29},
  8--10 (2004).

\bibitem{Ma2016NC}
L.~Ma, S.~Li, V.~M. Fomin, M.~Hentschel, J.~B. G{\"{o}}tte, Y.~Yin, M.~R.
  Jorgensen, and O.~G. Schmidt, \enquote{{Spin-orbit coupling of light in
  asymmetric microcavities.}} {\protect\JournalTitle{Nat. Commun.}} \textbf{7},
  10983 (2016).

\bibitem{NP2023}
J.~Wang, S.~Valligatla, Y.~Yin, L.~Schwarz, M.~Medina-S{\'a}nchez, S.~Baunack,
  C.~H. Lee, R.~Thomale, S.~Li, V.~M. Fomin, L.~Ma, and O.~G. Schmidt,
  \enquote{Experimental observation of berry phases in optical m{\"o}bius-strip
  microcavities,} {\protect\JournalTitle{Nat. Photonics}} \textbf{17}, 120--125
  (2023).

\bibitem{Strelow2012PRB}
C.~Strelow, C.~M. Schultz, H.~Rehberg, M.~Sauer, H.~Welsch, A.~Stemmann,
  C.~Heyn, D.~Heitmann, and T.~Kipp, \enquote{Light confinement and mode
  splitting in rolled-up semiconductor microtube bottle resonators,}
  {\protect\JournalTitle{Phys. Rev. B}} \textbf{85}, 155329 (2012).

\bibitem{SumetskyReview}
M.~Sumetsky, \enquote{Optical bottle microresonators,}
  {\protect\JournalTitle{Prog. Quantum Electron.}} \textbf{64}, 1--30 (2019).

\bibitem{Strelow2008PRL}
C.~Strelow, H.~Rehberg, C.~M. Schultz, H.~Welsch, C.~Heyn, D.~Heitmann, and
  T.~Kipp, \enquote{Optical microcavities formed by semiconductor microtubes
  using a bottlelike geometry,} {\protect\JournalTitle{Phys. Rev. Lett.}}
  \textbf{101}, 127403 (2008).

\bibitem{Li2013PRA}
S.~Li, L.~Ma, S.~Böttner, Y.~Mei, M.~R. Jorgensen, S.~Kiravittaya, and O.~G.
  Schmidt, \enquote{Angular position detection of single nanoparticles on
  rolled-up optical microcavities with lifted degeneracy,}
  {\protect\JournalTitle{Phys. Rev. A}} \textbf{88}, 033833 (2013).

\bibitem{Fang2016}
Y.~Fang, S.~Li, and Y.~Mei, \enquote{Modulation of high quality factors in
  rolled-up microcavities,} {\protect\JournalTitle{Phys. Rev. A}} \textbf{94},
  033804 (2016).

\bibitem{Tian2018}
Z.~Tian, S.~Li, S.~Kiravittaya, B.~Xu, S.~Tang, H.~Zhen, W.~Lu, and Y.~Mei,
  \enquote{Selected and enhanced single whispering-gallery mode emission from a
  mesostructured nanomembrane microcavity,} {\protect\JournalTitle{Nano Lett.}}
  \textbf{18}, 8035--8040 (2018).

\end{thebibliography}

\end{document}